\documentclass[conference]{IEEEtran}
\IEEEoverridecommandlockouts
\usepackage{cite}
\usepackage{amsmath,amssymb,amsfonts}
\usepackage{algorithmic}
\usepackage{graphicx}
\usepackage{subcaption}
\usepackage{textcomp}
\usepackage{xcolor}
\usepackage{lipsum}
\usepackage{csquotes}

\begin{document}

\title{RAN Functional Split Options for Integrated Terrestrial and Non-Terrestrial 6G Networks}

\author{\IEEEauthorblockN{ Mohamed Rihan, Tim Düe, MohammadAmin Vakilifard, Dirk Wübben, Armin Dekorsy}
\IEEEauthorblockA{\textit{Department of Communications Engineering},
\textit{University of Bremen},
Bremen, Germany. \\
\{elmeligy, duee, vakilifard, wuebben, dekorsy\}@ant.uni-bremen.de}}

\maketitle

\begin{abstract}
Leveraging non-terrestrial platforms in 6G networks holds immense significance as it opens up opportunities to expand network coverage, enhance connectivity, and support a wide range of innovative applications, including global-scale Internet of Things and ultra-high-definition content delivery. To accomplish the seamless integration between terrestrial and non-terrestrial networks, substantial changes in radio access network (RAN) architecture are required. These changes involve the development of new RAN solutions that can efficiently manage the diverse characteristics of both terrestrial and non-terrestrial components, ensuring smooth handovers, resource allocation, and quality of service across the integrated network ecosystem. Additionally, the establishment of robust interconnection and communication protocols between terrestrial and non-terrestrial elements will be pivotal to utilize the full potential of 6G technology. Additionally, innovative approaches have been introduced to split the functionalities within the RAN into centralized and distributed domains. These novel paradigms are designed to enhance RAN's flexibility while simultaneously lowering the costs associated with infrastructure deployment, all while ensuring that the quality of service for end-users remains unaffected. In this work, we provide an extensive examination of various Non-Terrestrial Networks (NTN) architectures and the necessary adaptations required on the existing 5G RAN architecture to align with the distinct attributes of NTN. Of particular significance, we emphasize the crucial RAN functional split choices essential for the seamless integration of terrestrial and non-terrestrial components within advanced 6G networks. 
\end{abstract}

\begin{IEEEkeywords}
Non-terrestrial networks (NTN), open - radio access network (O-RAN), Satellites, Sixth generation (6G), RAN functional splits.
\end{IEEEkeywords}

\section{Introduction}
Traditionally, satellite communications and terrestrial networks have been developed separately. However, with the advent of 5G, these two networks are increasingly being viewed as complementary technologies. 5G Non-Terrestrial Networks (NTN) can provide global coverage and high capacity, while terrestrial networks can provide low latency and high reliability \cite{R01,R02}. Accordingly, the 3rd Generation Partnership Project (3GPP) has worked on standardizing the implementation of such NTN and has already completed the first 5G New Radio (NR) specifications and is working on solutions to support NTN in 5G NR systems. Several projects, such as SAT5G, are also investigating the integration of 5G NTN with terrestrial networks \cite{R03,R04}.

One of the key requirements for the integration of 5G NTN with terrestrial networks is service continuity \cite{R05}. This means that users should be able to seamlessly handover between the two networks without experiencing any interruption in service. The integration of 5G NTN with terrestrial networks can offer a number of benefits, including global coverage, high capacity, and improved reliability. In the context of Global coverage, NTN can provide coverage in areas that are not well-served by terrestrial networks \cite{R06}. Regarding the high capacity requirement, NTN can provide high capacity for applications such as 5G Fixed Wireless Access and Internet of Things \cite{R07}. For the improved reliability requirement, NTN can improve the reliability of 5G networks by providing an alternative path for data traffic. The integration of 5G NTN with terrestrial networks is not easy to realize, but it has the potential to revolutionize the way we communicate \cite{R08,R09}.

On a parallel theme, as 5G applications and services continue to advance, novel RAN architectures and protocols are emerging as contenders to meet the evolving network demands. Among these contenders, network densification has been proposed as a means to augment network capacity significantly \cite{R10, R11, R12}. One transformative factor in this evolution is the integration of virtualization, which is reshaping communication networks and the architecture of RAN, including the repositioning of Radio Units (RUs) and Base Band Units (BBUs), traditionally housed within cellular Base Stations. The attempts toward virtualization have been initiated within 4G era by introducing the concept of IP Multimedia Subsystem (IMS) virtualization, which then extended to IMS cloudification \cite{R12b,R12c}. The 3GPP has introduced the concept of virtualizing network functions and functional splitting to advance RAN centralization while concurrently mitigating the overall costs associated with network densification efforts \cite{R01,R02}. Within this innovative RAN architecture, the functions traditionally housed in the 5G BBU are divided into distinct functional blocks, prominently featuring the Centralized Unit (CU), Distributed Unit (DU), and the RU. These components serve as the foundational elements of the Next Generation RAN (NG-RAN). The primary objective is to enable the deployment of flexible, cost-effective, energy-efficient, and straightforward Remote Radio Heads (RRHs). These RRHs offer a plethora of benefits, including the ability to jointly process radio signals, distribute network loads more effectively, extend network coverage, and optimize power consumption, thus enhancing the overall network performance and efficiency \cite{R13, R14, R15,R16}.

The remainder of this paper is organized as follows. \textbf{Section} \ref{section2} gives an overview on different aspects of NTN, including their definition, components, and architectures, with their potential service continuity and multi-connectivity capability. In \textbf{Section} \ref{section3}, we explore different RAN functional splits that have been proposed by different standardization bodies. A comprehensive overview of the open RAN architecture is explained in \textbf{Section} \ref{section4}. The potential functional split options for NTN are presented in \textbf{Section} \ref{section5}. Finally, concluding remarks are given in \textbf{Section} \ref{section6}. 

\section{Non-Terrestrial Networks}
\label{section2}
\subsection{NTN Definition}
NTN are wireless communication systems that operate above the Earth's surface, involving different space-borne and airborne platforms. Space-borne platforms are typically placed in orbit around the Earth \cite{R06, R07}. They can be further classified into three categories:
\begin{itemize}
\item \textbf{Geostationary Earth Orbiting (GEO) satellites} are located at an altitude of $35,786$ kilometers above the Earth's equator. They appear fixed in the sky to ground observers, because their orbital period matches the Earth's rotation period.
\item \textbf{Medium Earth Orbiting (MEO) satellites} are located at an altitude of $7,000$ to $25,000$ kilometers above the Earth. They have a smaller beam footprint than GEO satellites, which means they can provide coverage to a smaller area.
\item \textbf{Low Earth Orbiting (LEO) satellites} are located at an altitude of $300$ to $1,500$ kilometers above the Earth. They have the smallest beam footprint of the three categories, which means they can provide coverage to a very small area.
\end{itemize}

\noindent Airborne platforms are typically placed at an altitude of $8$ to $50$ kilometers above the Earth \cite{R08}. They can be further classified into two categories:
\begin{itemize}
\item  \textbf{Unmanned Aircraft Systems (UAS)}: UAS are typically used for military or commercial applications. They can be used to provide temporary or emergency coverage in areas where terrestrial networks are not available.
\item \textbf{High Altitude Platform Systems (HAPS)}: HAPS are typically used for civilian applications, such as providing broadband internet access to rural areas. They have a longer lifespan than UAS and can provide more reliable coverage.
\end{itemize}

\noindent In addition to the platform, the NTN access also includes the following components \cite{R06,R07}:
\begin{itemize}
\item \textbf{NTN terminal}: The NTN terminal is the device that connects the user to the NTN network. It can be a 3GPP User Equipment (UE) or a specific satellite terminal.
\item \textbf{NTN gateway}: The NTN gateway is a logical node that connects the NTN platform to the 5G core network.
\item \textbf{Service link}: The service link is the radio link between the NTN terminal and the NTN platform.
\item \textbf{Feeder link}: The feeder link is the radio link between the NTN gateway and the NTN platform.
\end{itemize}
\noindent The choice of NTN platform and other components depends on the specific application requirements, such as the required coverage area, the data rate, and the latency.

\subsection{NTN Architectures}
\begin{figure}[t]
    \centering
    \begin{subfigure}{0.45\textwidth}
        \centering
        \includegraphics[width=0.95\linewidth]{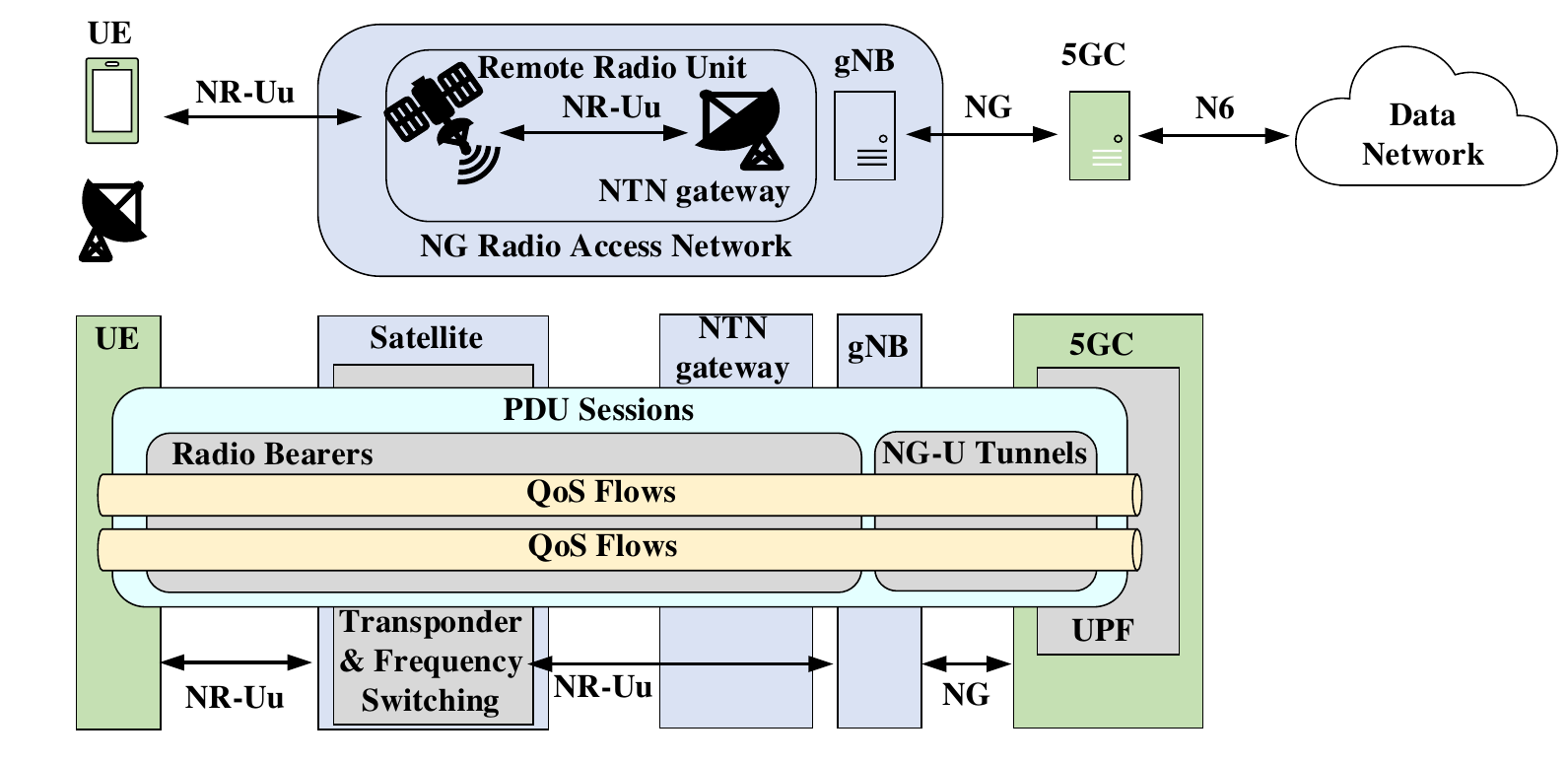}
        \caption{}
        \label{fig1a}
    \end{subfigure}
    \hfill
    \begin{subfigure}{0.45\textwidth}
        \centering
        \includegraphics[width=0.95\linewidth]{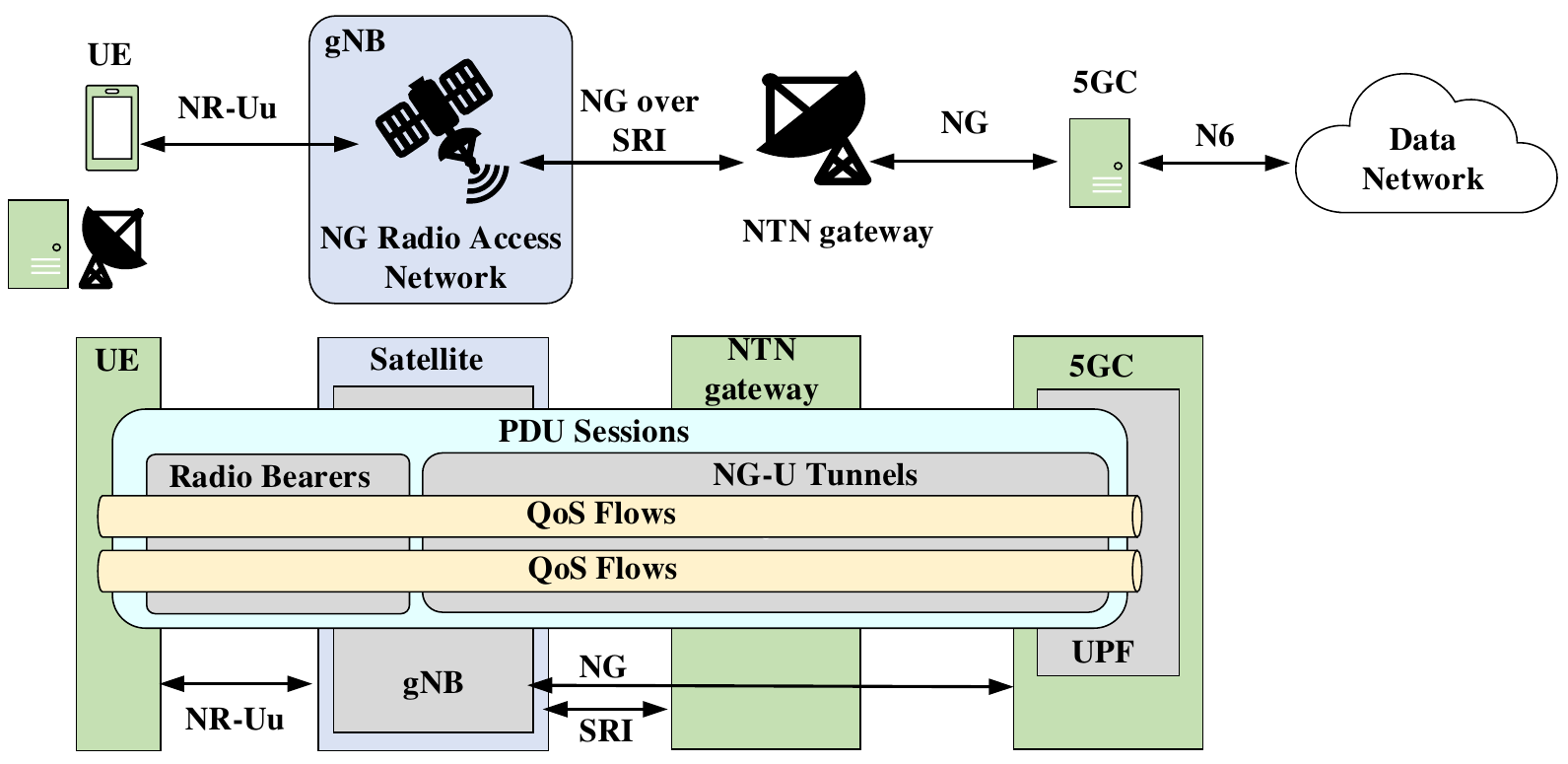}
        \caption{}
        \label{fig1b}
    \end{subfigure}
    \hfill
    \begin{subfigure}{0.45\textwidth}
        \centering
        \includegraphics[width=0.95\linewidth]{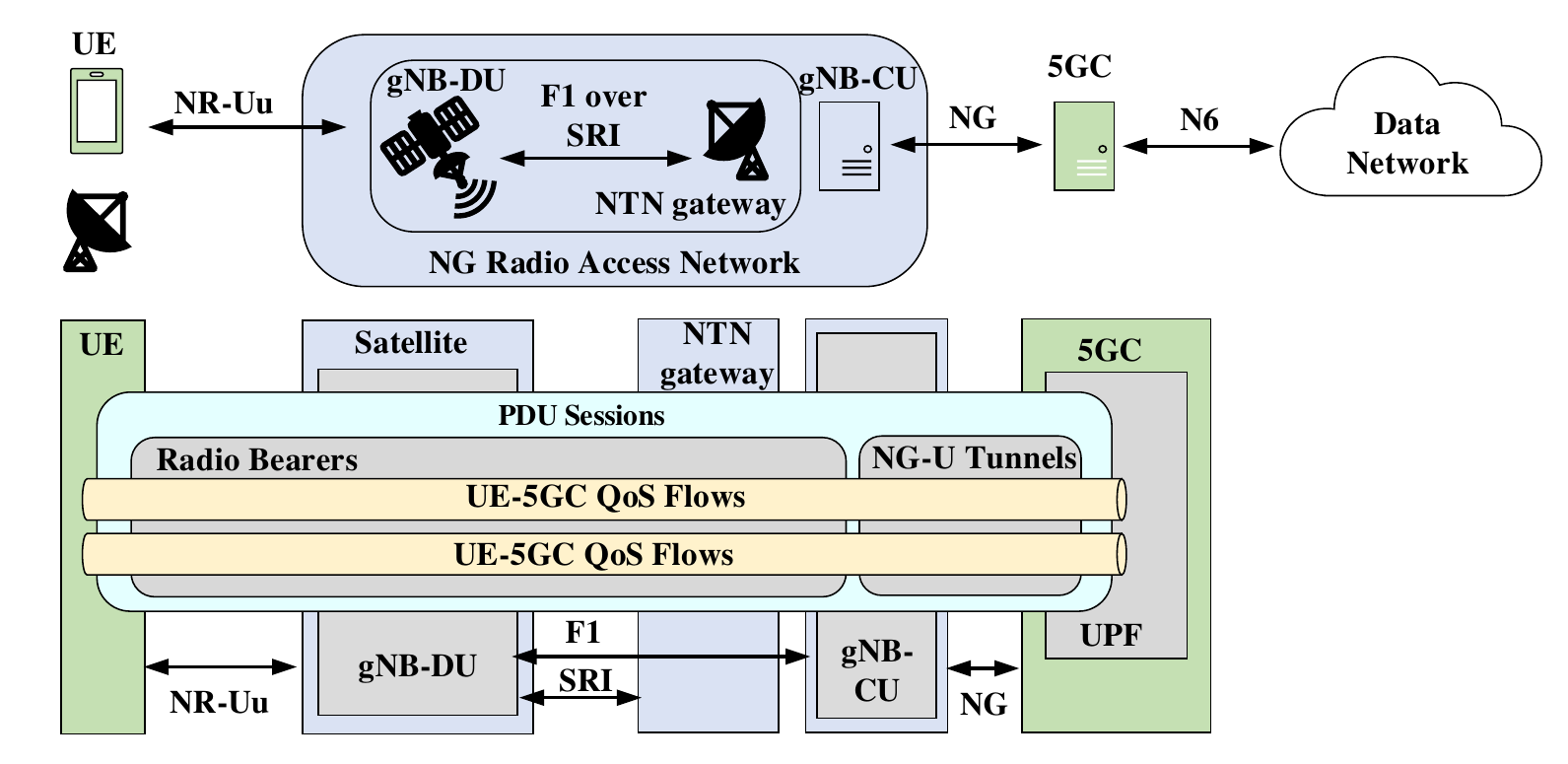}
        \caption{}
        \label{fig1c}
    \end{subfigure}
    \caption{Satellite access architectures: (a) Transparent payload-based satellite. (b) Regenerative payload-based satellite. (c) Regenerative satellite-based gNB-DU.}
    \label{fig1}
    \vspace{-0.5cm}
\end{figure}
In order to integrate satellites into 5G networks, new interfaces and protocols are being developed. These new interfaces will allow satellites to act as gNBs (NR logical node), or ground-based network nodes. This means that satellites will be able to directly communicate with 5G devices, such as smartphones and laptops \cite{R17}. The 3GPP TR 38.821 has defined NTN reference scenarios for both transparent and regenerative satellite with either GEO and LEO satellites \cite{R06}.

There are two main types of satellite-based NG-RAN architectures: \textbf{transparent} and \textbf{regenerative}. \textbf{Transparent architectures} are the simplest type. The satellite simply relays the signal from the 5G core network to the 5G device, and vice versa. This is similar to how a TV satellite works \cite{R04}. \textbf{Regenerative architectures} are more complex. The satellite has its own processing capabilities, so it can regenerate the signal before sending it to the 5G device. This enables better performance, such as higher reliability and higher data rates. There are also two main types of satellite access architectures: satellite access and relay-like. \textbf{Satellite access architectures} connect the 5G device directly to the satellite. This is the most common type of satellite access architecture \cite{R06}. \textbf{Relay-like architectures} use a relay node to connect the 5G device and the satellite. This can be useful in areas where the satellite signal is not strong enough. The choice of satellite architecture depends on the specific application requirements. For example, transparent architectures are a good choice for applications that require low latency, while regenerative architectures are a good choice for applications that require high data rates\cite{R11,R12}.

\subsubsection{Satellite access architectures}
Figure \ref{fig1} shows three different satellite access architectures for 5G networks \cite{R01,R02, R06}.

\begin{figure}
    \centering
    \begin{subfigure}{0.5\textwidth}
        \centering
        \includegraphics[width=\linewidth]{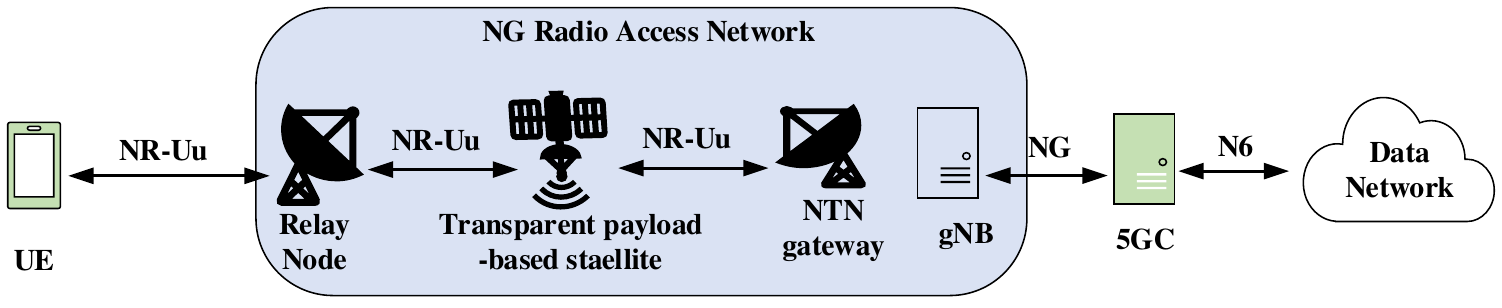}
        \caption{}
        \label{fig2a}
    \end{subfigure}
    \hfill
    \begin{subfigure}{0.5\textwidth}
        \centering
        \includegraphics[width=\linewidth]{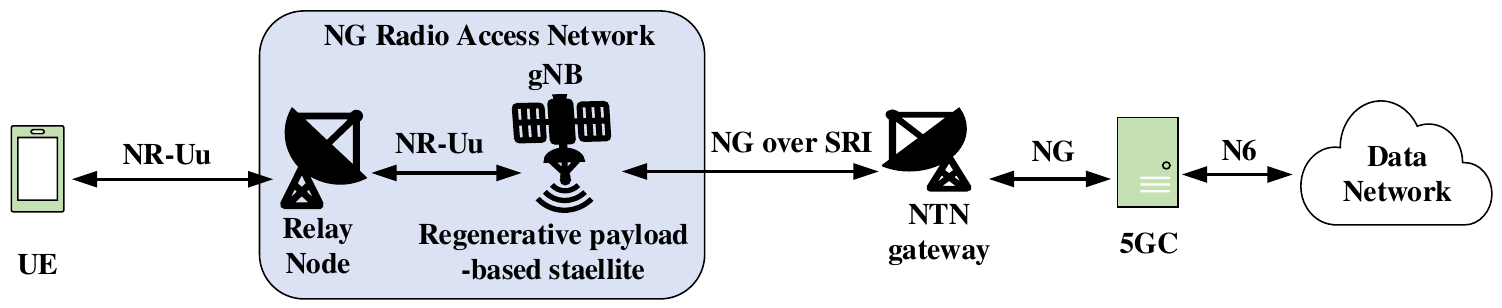}
        \caption{}
        \label{fig2b}
    \end{subfigure}
    \hfill
    \begin{subfigure}{0.5\textwidth}
        \centering
        \includegraphics[width=\linewidth]{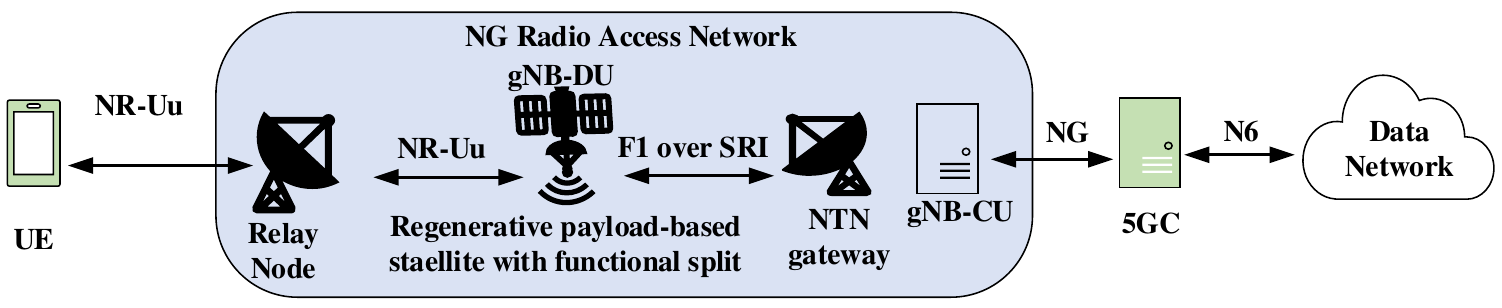}
        \caption{}
        \label{fig2c}
    \end{subfigure}
    \caption{Relay-like architectures: (a) Transparent payload-based satellite. (b) Regenerative payload-based satellite. (c) Regenerative satellite-based gNB-DU.}
    \label{fig2}
    \vspace{-0.5cm}
\end{figure}

\noindent \textbf{Transparent architecture (Figure \ref{fig1a})}: The NTN platform simply relays the signal from the NTN gateway to the NTN terminal and vice versa. The satellite radio interface (SRI) on the feeder link is the same as the radio interface on the service link (i.e., NR-Uu). The NTN gateway can forward the NR signal of the NR-Uu interface to the gNB. One or more transparent satellites can be connected to the same gNB on the ground \cite{R06,R07}.

\noindent \textbf{Regenerative architecture (Figure \ref{fig1b})}: The NTN platform has its own processing capabilities, so it can regenerate the signal before sending it to the NTN terminal. This allows for better performance, such as lower latency and higher data rates. The NR-Uu interface is on the service link between the NTN terminal and the NTN platform. The radio interface between the NTN platform and the 5G Core Network (5GC) is NG, which is over SRI in the air path between the NTN platform and the NTN-gateway. Inter-Satellite Links are transport links between the NTN platforms \cite{R01, R02, R06,R07}.

\noindent \textbf{5G NR-friendly architecture (Figure \ref{fig1c})}: This architecture is based on the regenerative architecture, but the NTN platform also acts as a gNB-DU. This means that the NTN platform can take on some of the processing tasks that are normally performed by the gNB on the ground. This can improve the performance of the network, especially in areas with high traffic demand \cite{R10,R11, R12}.

\subsubsection{Relay-like architectures}
Figure \ref{fig2} shows three different relay-like architectures for 5G networks \cite{R06, R07}.

\noindent \textbf{Transparent architecture (Figure \ref{fig2a})}: The access network forwards the NR signal to the NTN terminal through a relay node, which receives it from the transparent payload-based satellite. The relay node simply relays the signal without any processing.

\noindent \textbf{Regenerative architecture with full gNB (Figure \ref{fig2b})}: The regenerative payload-based satellite includes a full gNB. This means that the satellite can perform all of the processing tasks that are normally performed by the gNB on the ground. The relay node forwards the NR signal received from the satellite to the NTN terminal.

\noindent \textbf{Regenerative architecture with partial gNB (Figure \ref{fig2c})}: The regenerative payload-based satellite includes a partial gNB. This means that the satellite can only perform some of the processing tasks that are normally performed by the gNB on the ground. The relay node forwards the NR signal received from the satellite to the NTN terminal, and the NTN terminal performs the remaining processing tasks.

\subsection{Service Continuity and Multi-Connectivity}
The seamless integration of NTN with terrestrial networks holds utmost importance in ensuring the uninterrupted delivery of services and scalability in the era of 5G and beyond. This integrated terrestrial-NTN system brings forth a multitude of advantages, not only in densely populated urban areas, but also in more remote rural regions, meeting the ambitious performance targets set by 5G in terms of data rates and reliability. It guarantees robust connectivity within bustling environments like concert venues, sports stadiums, city centers, and shopping malls, as well as for users in motion, whether they are passengers on high-speed trains, travelers on airplanes, or cruise ship passengers \cite{R01, R02}.

However, the quest for service continuity in 5G systems extends beyond the seamless handover between terrestrial NG-RAN and NTN NG-RAN. It also encompasses the need for uninterrupted connectivity between two NTN NG-RANs. To address this requirement, 3GPP's Technical Report 38.821 investigates the concept of multi-connectivity, enabling simultaneous access to both NTN and terrestrial NG-RANs, or even between two distinct NTN NG-RANs. Consequently, this prompts the exploration of architectures that can effectively support such multi-connectivity scenarios \cite{R06, R07}.

\begin{figure*}[ht]
    \centering
    \begin{subfigure}{0.49\textwidth}
    \centering    \includegraphics[width=1.0\textwidth,height=0.5\textwidth]{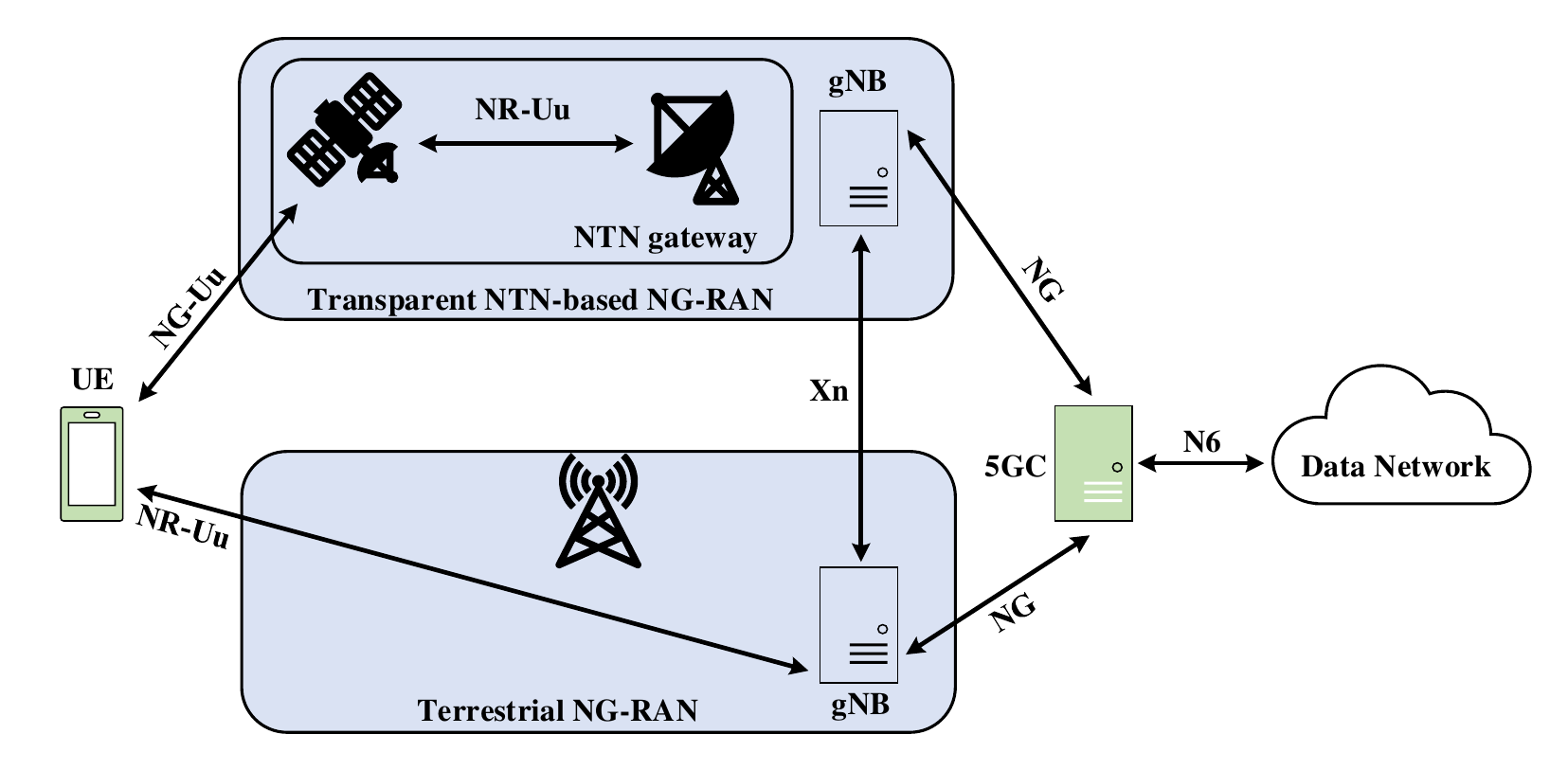}
    \caption{A transparent NTN-based NG-RAN and a terrestrial NG-RAN.}
    \label{fig3a}
    \end{subfigure}
    \hfill
    \begin{subfigure}{0.49\textwidth}
    \centering
    \includegraphics[width=1.0\textwidth,height=0.5\textwidth]{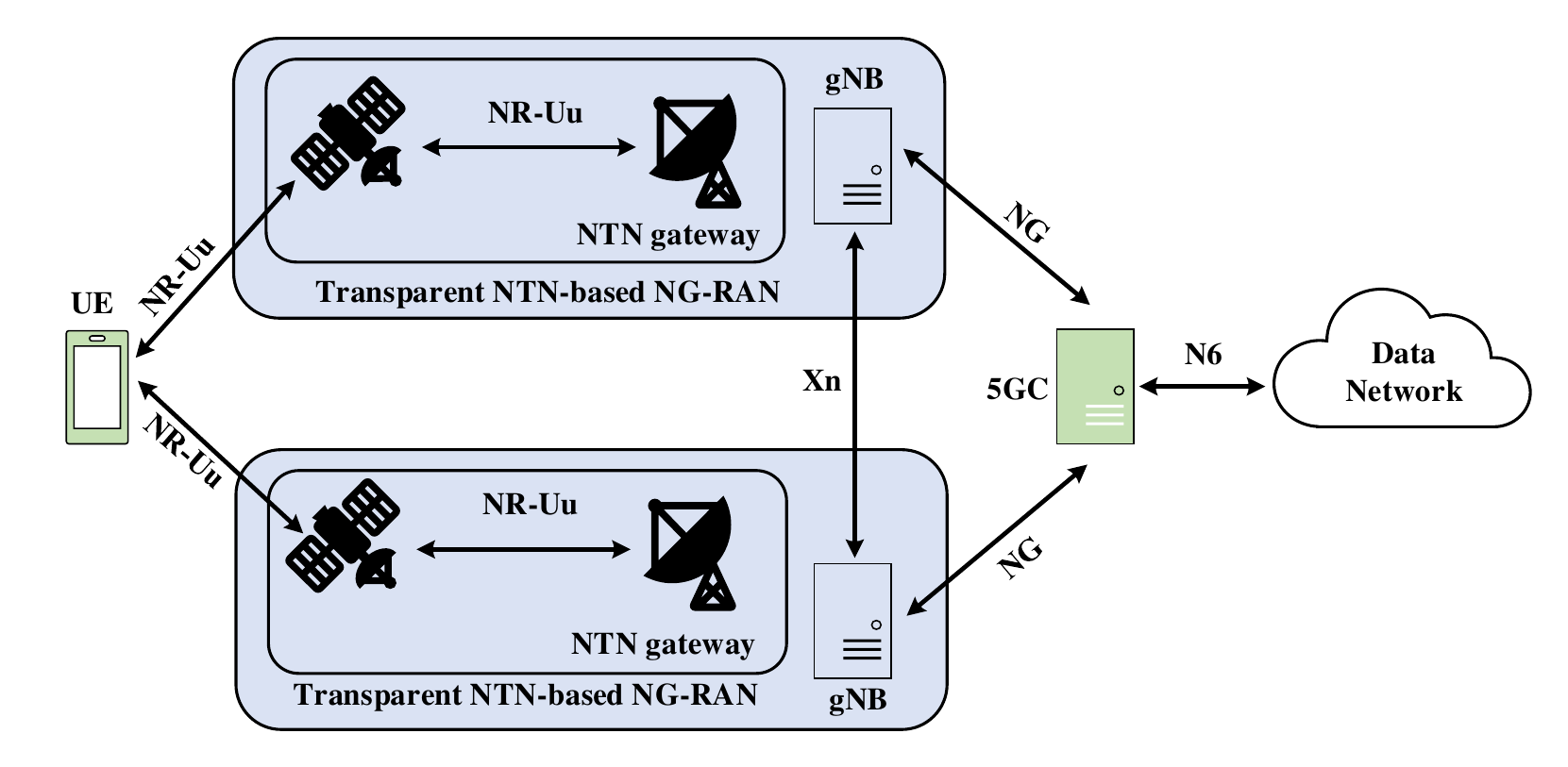}
    \caption{Two transparent NTN-based RANs.}
    \label{fig3b}
    \end{subfigure}
    \begin{subfigure}{0.49\textwidth}
    \centering
    \includegraphics[width=1.0\textwidth,height=0.5\textwidth]{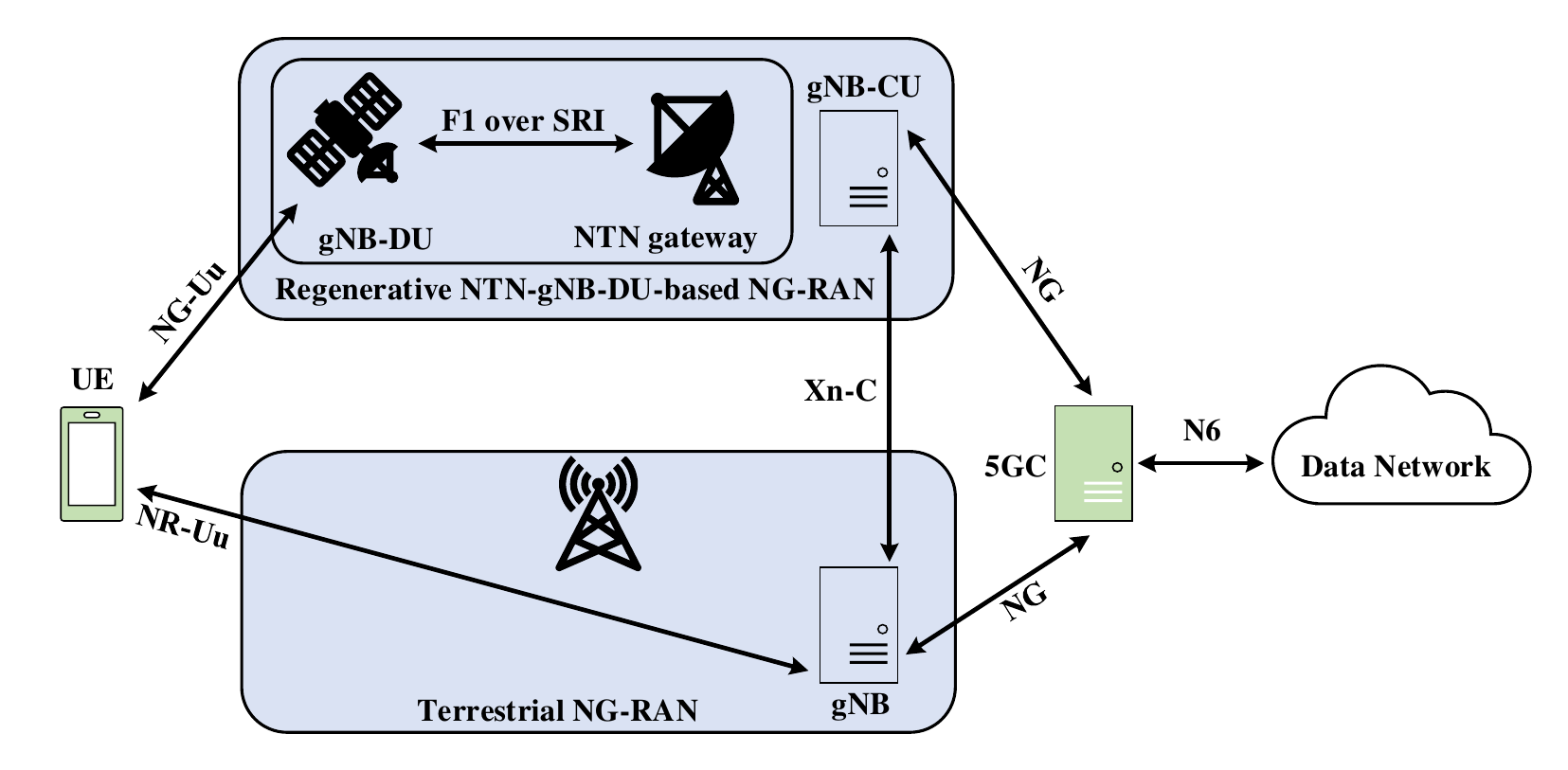}
    \caption{A regenerative NTN-gNB-DU-based NG-RAN and a terrestrial NG-RAN.}
    \label{fig3c}
    \end{subfigure}
    \hfill
    \begin{subfigure}{0.49\textwidth}
    \centering
    \includegraphics[width=1.0\textwidth,height=0.5\textwidth]{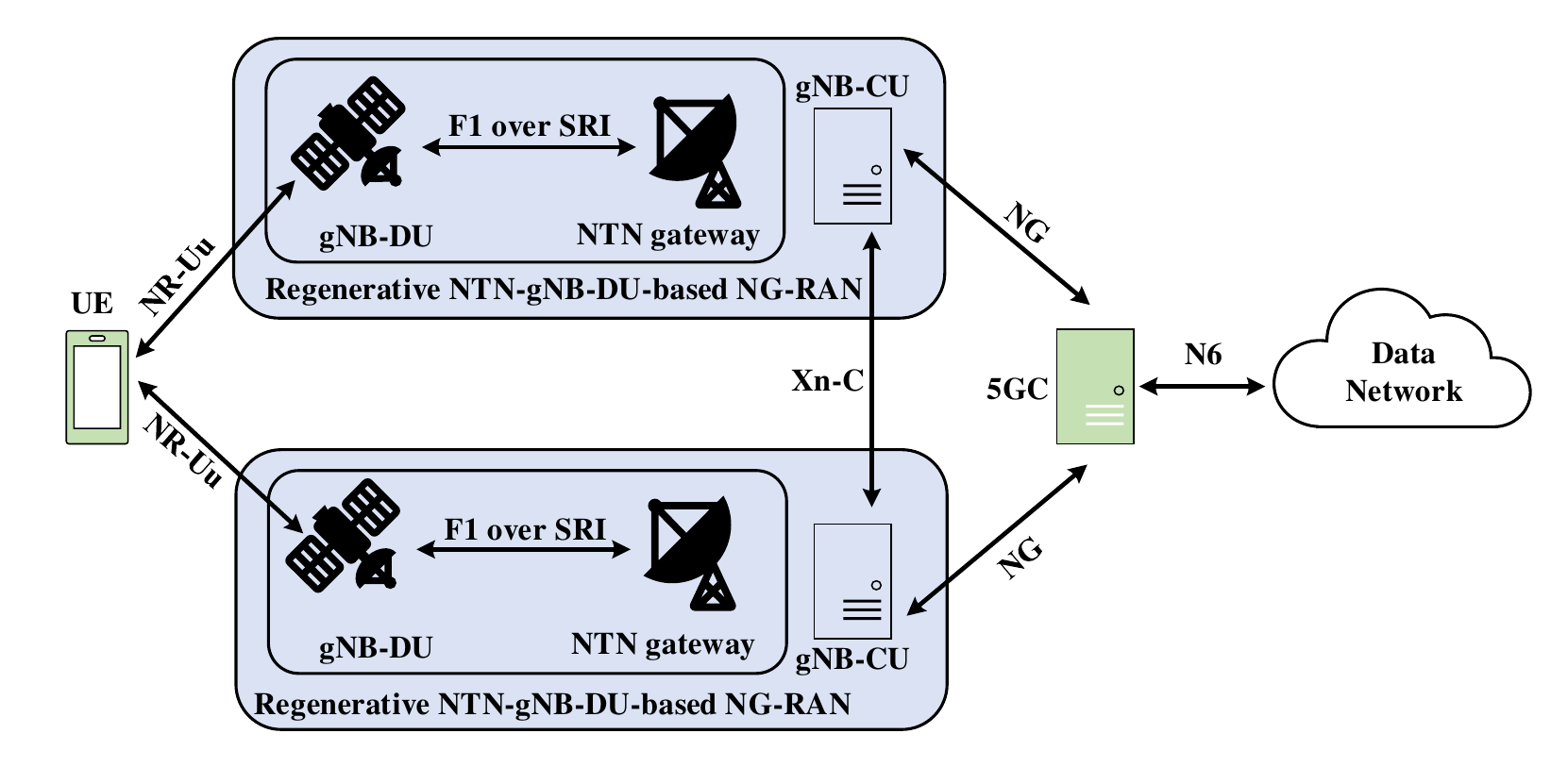}
    \caption{Two regenerative NTN-gNB-DU-based RANs.}
    \label{fig3d}
    \end{subfigure}
    \begin{subfigure}{0.49\textwidth}
    \centering
    \includegraphics[width=1.0\textwidth,height=0.5\textwidth]{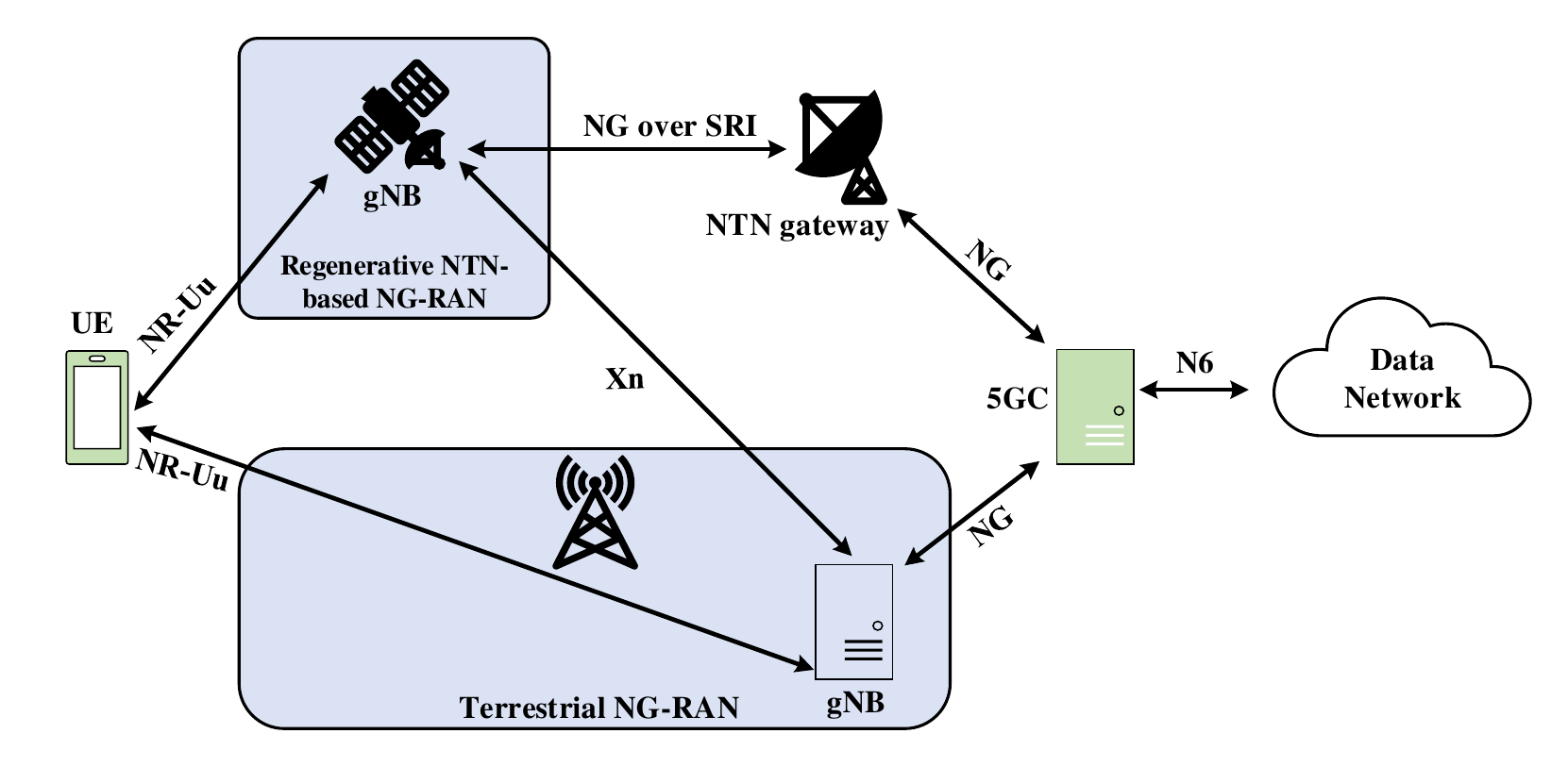}
    \caption{A regenerative NTN-based NG-RAN and a terrestrial NG-RAN.}
    \label{fig3e}
    \end{subfigure}
    \hfill
    \begin{subfigure}{0.49\textwidth}
    \centering
    \includegraphics[width=1.0\textwidth,height=0.5\textwidth]{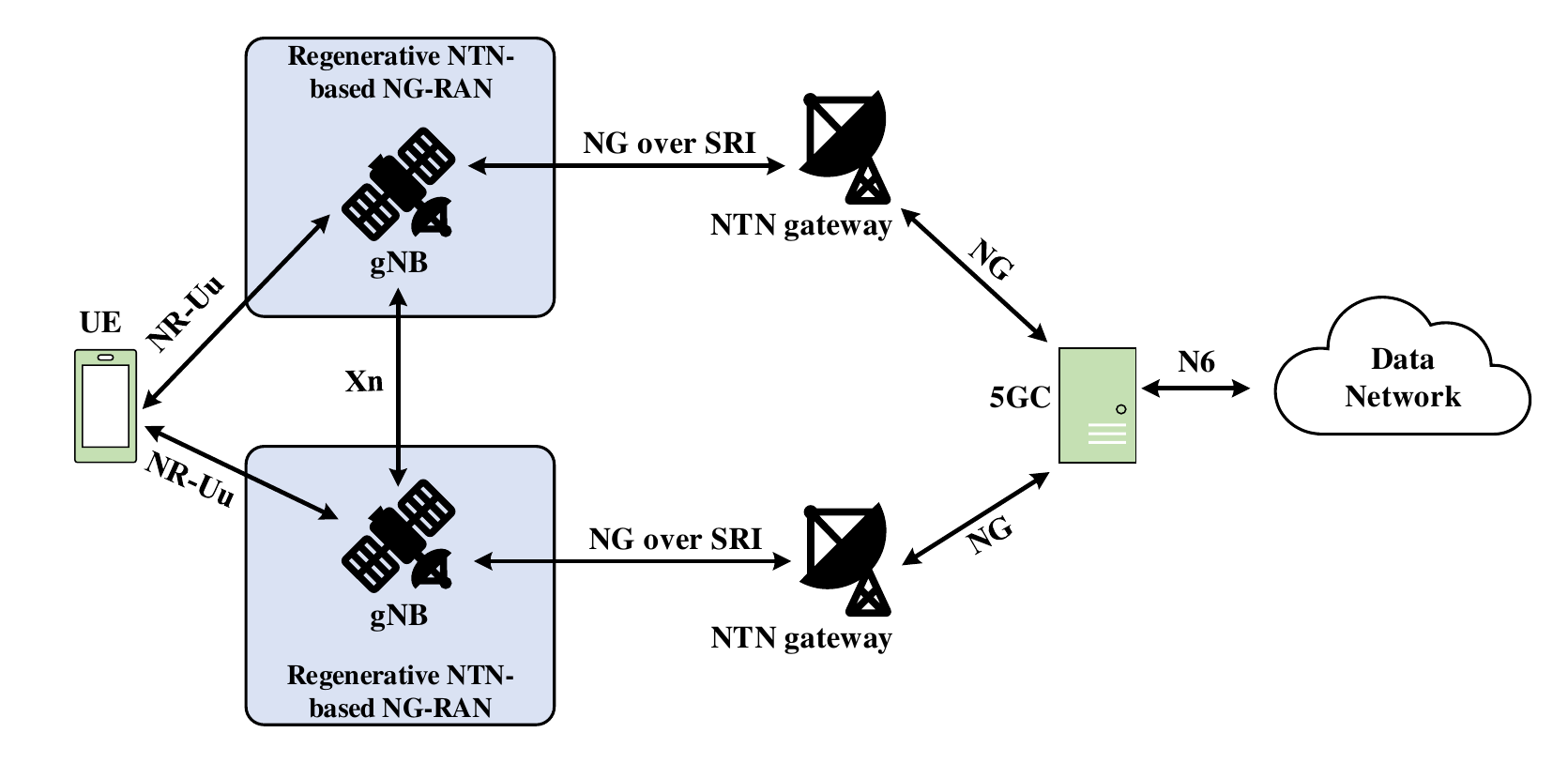}
    \caption{Two regenerative NTN-based RANs.}
    \label{fig3f}
    \end{subfigure}
    \caption{RAN architectures supporting multi-connectivity.}
    \label{fig3}
\end{figure*} 

Figure \ref{fig3} shows the different multi-connectivity architectures for NTN. In Figure \ref{fig3a}, the ground terminal is connected simultaneously to the 5GC via both a transparent NTN-based NG-RAN and a terrestrial NG-RAN. The NTN gateway is located in the Public Land Mobile Network area of the terrestrial NG-RAN. In Figure \ref{fig3b}, this architecture combines two transparent NTN-based NG-RANs, which can be either GEO or LEO satellites, or a combination of both. This scenario can be used to provide services to users in unserved areas. LEO satellites are used to deliver delay-sensitive traffic because they have lower propagation delay than GEO satellites. GEO satellites are used to provide additional bandwidth, higher throughput, reliability due to the large covered area and lack of frequent hand overs. In Figure \ref{fig3c}, this architecture combines a regenerative NTN-gNB-DU-based NG-RAN and a terrestrial NG-RAN. The functional split is applied in this type of architecture, which means that the NTN platform acts as a distributed unit of the gNB, with the central unit located on the ground. This scenario can be used to provide services to users in under-served areas. Multi-connectivity can also involve two regenerative NTN-gNB-DU-based NG-RANs (see Figure \ref{fig3d}). In Figure \ref{fig3e}, this architecture combines two regenerative NTN-based NG-RANs, which can be either GEO or LEO satellites, or a combination of both. The NTN platforms in this architecture perform all the gNB tasks, meaning that the functional split is not applied. Multi-connectivity can also involve a regenerative NTN-based NG-RAN and a terrestrial NG-RAN (see Figure \ref{fig3f}). 

\section{General RAN Functional Split Options of 5G networks}
\label{section3}
\begin{figure}
\centering
\includegraphics[width=0.9\linewidth]{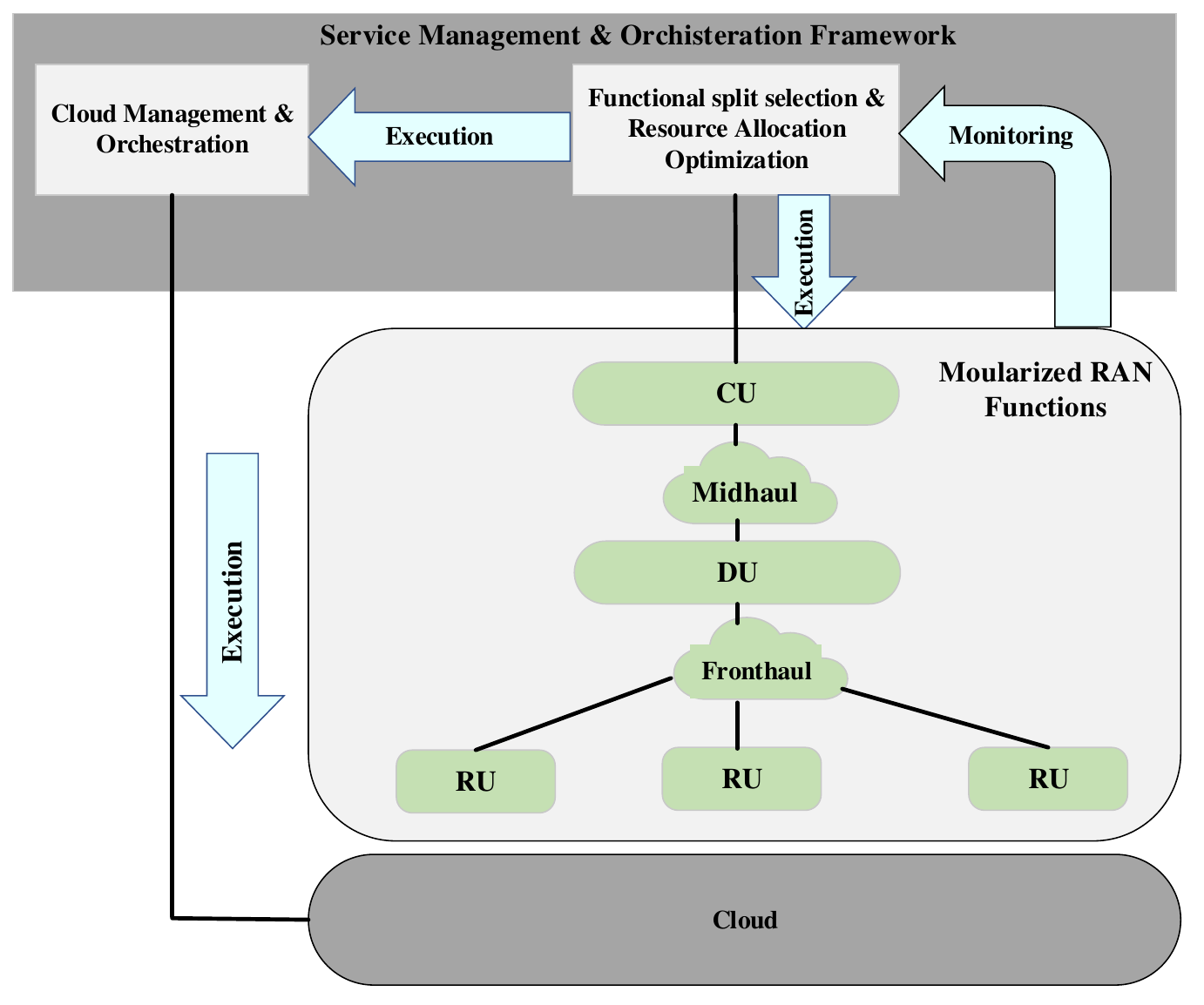}
\caption{Modular Flexible RAN architecture.}
\label{fig4}
\vspace{-0.5cm}
\end{figure}

Previous generations of RAN architectures, including 2G, 3G, and 4G, were characterized by their monolithic building blocks, with limited interactions between logical nodes. However, as the NR concept emerged, there was a growing recognition that dividing the gNB into Central Units (CUs) and Distributed Units (DUs) could introduce much-needed flexibility, as indicated in the modular flexible RAN architecture shown in Figure \ref{fig4} \cite{R17, R18, R19}. CU and DU are connected through an integrated fronthaul/backhaul network, often referred to as X-haul. The adoption of flexible hardware and software implementations offered the potential for scalable and cost-effective network deployments, provided that these components could seamlessly inter-operate and be sourced from different vendors \cite{R20}. This split architecture, differentiating between central and distributed units, opens up avenues for improved coordination of performance features, efficient load management, real-time performance optimization, and the ability to adapt to diverse use cases and Quality of Service (QoS) requirements \cite{R21}. These use cases encompass a wide range of applications, such as gaming, voice, and video, each with varying degrees of latency tolerance and transport dependencies. Furthermore, this adaptability extends to different deployment scenarios, whether they be in rural or urban environments, each with distinct access to transport infrastructure, such as fiber-optic networks \cite{R22}.

\begin{figure}
\centering
\includegraphics[width=0.85\linewidth]{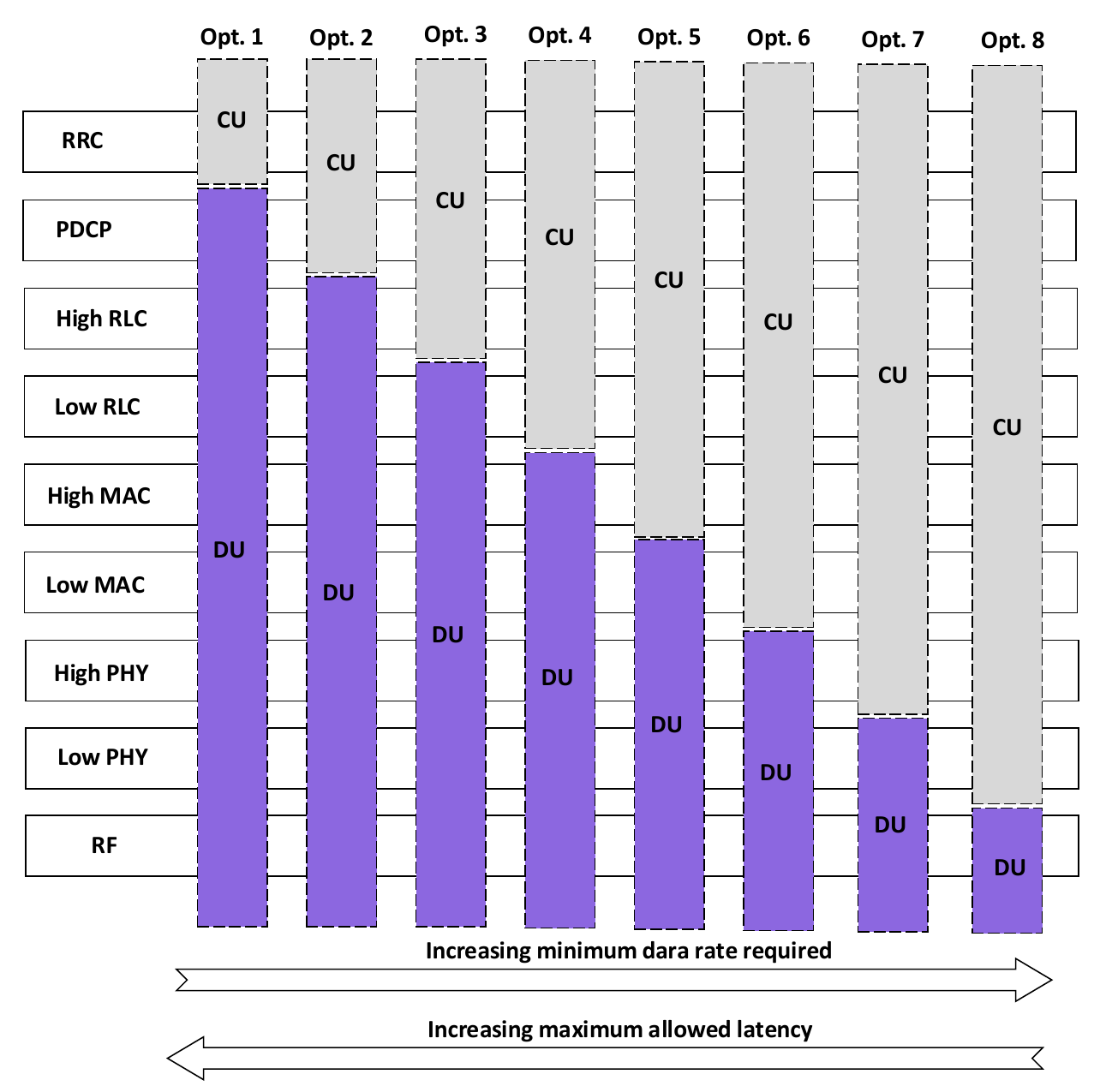}
\caption{Functional split options defined by 3GPP and allocation of protocol stack layers in the DU and CU.}
\label{fig5}
\end{figure}

The gNB's operation is conceptually structured as a series of functions, with the distribution of these functions being defined by functional splits. Within the 3GPP framework, there exist eight distinct functional split options, and Figure \ref{fig5} visually illustrates how these functions are allocated between the CU and DU. As we observe a higher degree of centralization, there emerges a notable improvement in resource management efficiency, leading to reduced DU complexity and associated costs. However, this comes at the expense of imposing more stringent data rate and latency requirements on the X-haul network \cite{R23, R24}. For instance, in the case of option 8, a demanding X-haul data rate of approximately 2.5 Gb/s is mandated for a 20 MHz bandwidth, with a maximum latency tolerance of 250 $\mu$s. Conversely, option 2 sets more relaxed data rate expectations, with 150 Mb/s in the downlink and 50 Mb/s in the uplink, alongside latency in the range of tens of milliseconds. The specific placement of each protocol layer plays a pivotal role in determining the functional split's characteristics. When Packet Data Convergence Protocol (PDCP) resides in the DU (as in option 1), a clear separation between control and user planes is established, making it an apt choice for scenarios involving edge computing, particularly relevant to Ultra-Reliable Low Latency Communication applications. As we move PDCP towards the CU (options 2 to 8), we enable centralized aggregation of traffic from both 5G and Long Term Evolution — Advanced. Within this context, the Radio Link Control (RLC) layer assumes responsibility for ARQ mechanisms. Centralizing ARQ within the CU contributes to enhancing X-haul reliability while concurrently reducing buffering and computational requirements in the DU \cite{R01, R25, R26}. This configuration aligns with options ranging from 3-1 (a sub-option of 3) to 8. The allocation of high MAC functions significantly influences the centralization or decentralization of the scheduler. Options 5 to 8 embrace centralized scheduling within the CU, a strategy well-suited for inter-cell coordination, albeit imposing stringent X-haul latency constraints. The lower-layer splits are defined by options 7 and 8, which, in turn, branch into three distinct PHY split variants, labeled as sub-options 7-1, 7-2, and 7-3 \cite{R01,R02}. In the case of option 8, an all-encompassing centralization occurs in the CU, except for RF components. This approach offers advantages such as the isolation of RF elements, facilitating PHY upgrades, the reuse of RF components, and the optimization of resource management. Amidst this spectrum of functional splits, certain options emerge as representative choices. For highly decentralized applications devoid of stringent cell coordination, 3GPP suggests option 2, particularly suitable for scenarios characterized by limited X-haul bandwidth and latency requirements. On the other hand, the Small Cell Forum (SCF) advocates for option 6, positioning it as the optimal split for cost-effective, low-capacity deployments. Meanwhile, the O-RAN Alliance lends its support to option 7-2, specifically the 7-2x variant, tailored for networks with high-capacity demands and stringent reliability criteria \cite{R25, R26}.

\section{Open RAN Architecture}
\label{section4}
\begin{figure}
\centering
\includegraphics[width=0.85\linewidth]{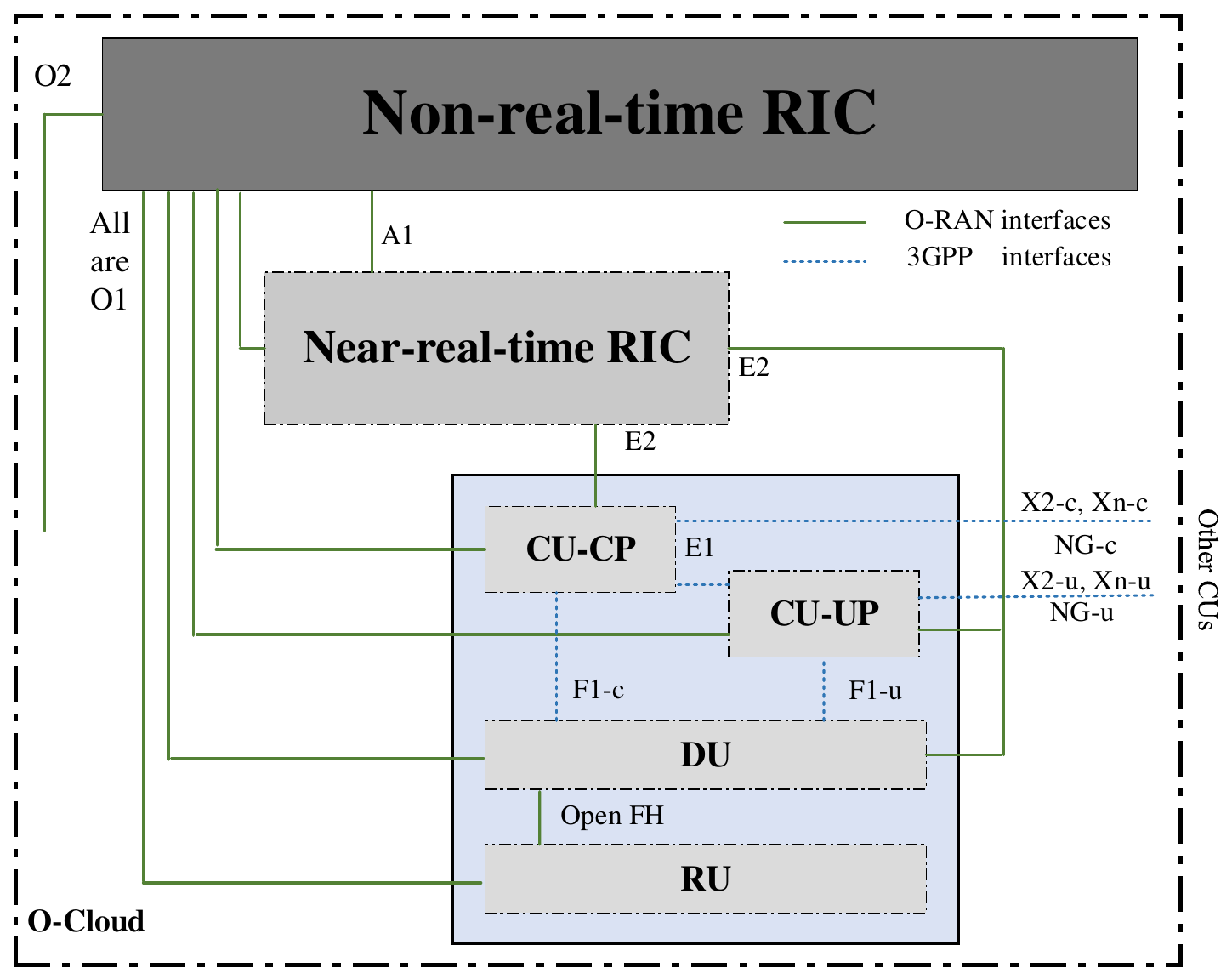}
\caption{High level O-RAN architecture.}
\label{fig6}
\end{figure}

Traditional mobile networks are monolithic, meaning that all the network functions are implemented in a single, proprietary hardware. This makes the networks difficult to reconfigure and inefficient to operate. Next-generation networks will use open RAN (O-RAN) architectures, which disaggregate the network functions into separate components that can be implemented on different hardware, see Figure \ref{fig6}. This will make the networks more flexible and efficient, and it will also open up new opportunities for innovation \cite{R34}.

\noindent The key concepts of open RAN are:
\begin{itemize}
\item \textbf{Disaggregation}: The network functions are split into separate components, such as the RU, the DU, and the CU. This makes it possible to use different vendors for different components, which can lead to lower costs and increased flexibility.
\item \textbf{Virtualization}: The network functions can be run on virtual machines, which makes it easier to deploy and manage the network.
\item \textbf{RAN Intelligent Controllers (RICs)}: RICs are software-based controllers that manage the RAN functions. They can be used to optimize the network performance and to introduce new features and services, this is consistent with the 5G America white paper that agreed on the fact that  one of the main pillars of O-RAN is taking advantage of AI/ML as Based on ongoing network and UE performance monitoring, the goal targets for optimization are programmatically created and modified using AI/ML-driven declarative rules.
\item \textbf{Open interfaces}: The open interfaces between the different components make it possible to interoperate with different vendors' equipment. This will help to create a more competitive and innovative RAN market.
\end{itemize}

\section{RAN Functional Split Options for NTN}
\label{section5}

In the NTN, the RAN functions are divided between the ground segment and the satellite's payload segment. The prescribed functional split architecture in \textbf{Section} \ref{section3}, as outlined in TR 38.821, places the CU on the ground, while the DUs and RUs are located aboard the satellite. The interface between CU and DU operates via the Space-to-Ground link and Inter-Node Links. In a static setup, the optimal functional split can be established during the design phase. However, the NTN's architecture is dynamic and undergoes frequent morphological changes \cite{R28, R29}. Consequently, the functional split must adapt dynamically to these shifts. To address this challenge, the NTN architecture, based on O-RAN principles explained in \textbf{Section} \ref{section4}, employs near-real-time Radio Intelligent Controllers (RT-RICs) to enable system-aware and proactive optimization of the functional split. The RICs gather network status data, leveraging it to compute the most efficient functional split and subsequently deploying network functions within CU and DU accordingly. A primary objective of this optimization is to minimize energy consumption within the satellite's payload, as the available power is limited. The RICs consider various factors, including user traffic type and volume, payload's computational capabilities, available power at any given moment, and the throughput and latency of the CU-DU physical feeder link \cite{R30, R31, R32, R33, R34}.

To ensure optimal operation, the RICs must possess the ability to forecast future network behavior and requirements, necessitating the use of AI algorithms. In addition to energy conservation, RICs can optimize the functional split to maximize the utilization of the feeder link, accounting for the time-varying performance characteristics of the link. The most challenging aspect of implementing a dynamic functional split lies in the high level of flexibility required for the satellite's payload. This can be achieved either through general-purpose computing processors, or by implementing individual RAN functions on specialized and isolated hardware, which can be activated independently. Presently, the general-purpose computing technology available consumes too much power to be used in payloads operating in Non-Geostationary Orbit. Conversely, the specialized hardware approach is complex and often under-utilizes the payload hardware. As a result, the current applicability of a dynamic functional split is limited to low-capacity services. However, with future developments in low-power, high-performance processors, the potential for its applicability can be expanded to encompass a wider range of service types .

\section{Conclusions}
\label{section6}
In this paper, we investigated the importance of functional split concept for the implementation of O-RAN infrastructure in NTN 6G networks. We first provided an overview of the different platforms that can be used to build NTN, including space-borne and airborne platforms. We then discussed the different use cases that can be supported by NTN, and the challenges that need to be addressed, in order to realize these use cases. We also introduced the concepts of service continuity and multi-connectivity in NTN, and explained how they can be used to improve the quality of service for users. We then identified the RAN functional split options proposed by 3GPP for 5G networks, and discussed the trade-offs between these options. We also reviewed the state-of-the-art works on optimizing the RAN functional split options in NTN. Our paper concludes that the RAN functional split is a key enabler for the implementation of O-RAN infrastructure in NTN 6G networks. The optimal RAN functional split for a particular NTN deployment will depend on a number of factors, such as the use cases that need to be supported, the available resources, and the cost constraints. However, the RAN functional split is a powerful tool that can be used to improve the performance, efficiency, and flexibility of NTN. 

\end{document}